\RequirePackage{fixltx2e}

\documentclass[aip,rsi,reprint]{revtex4-1} 

\usepackage{amsmath,amssymb,bm}
\usepackage{graphicx}\graphicspath{{}{graph/}{../graph/}}

\usepackage{ifpdf,ifluatex,ifxetex}
\newif\ifunicode
\ifxetex\unicodetrue\else\ifluatex\unicodetrue\fi\fi

\ifunicode

    \usepackage[english]{babel}

    \usepackage{fontspec}
    \defaultfontfeatures{Ligatures={TeX}}
    \usepackage[math-style=ISO,bold-style=ISO]{unicode-math}
    \mathversion{cambria}

\else

    \ifpdf
        \usepackage{cmap}
    \fi
    \usepackage[T1]{fontenc}
    \usepackage[utf8]{inputenc}
    \usepackage[english]{babel}
    \usepackage{csquotes}
    \usepackage{txfonts}
\fi

\usepackage[
    colorlinks
        ,linkcolor=violet
        ,filecolor=purple
        ,citecolor=teal
    ,unicode
    ,bookmarksnumbered
    ,pdfpagemode=UseOutlines
    ,pdfdisplaydoctitle=true
    ,bookmarksopenlevel = 1
    ,pdfusetitle
]{hyperref}
\usepackage{bookmark}

\DeclareMathOperator{\re}{Re}

\DeclareMathOperator{\const}{const}

\newcommand{\parder}[2]{\frac{\partial #1}{\partial #2}}
\newcommand{\tparder}[2]{{\partial #1}/{\partial #2}}
\newcommand{\dif}[1][]{\mathop{}\!\mathrm{d}
    \if\relax\detokenize{#1}\relax\else
        ^{\mkern-1.mu#1}\mkern-2.5mu
    \fi}

    \usepackage{xcolor}
    \definecolor{darkblue}{cmyk}{1.00, 0.50, 0.00, 0.40}

\begin{document}


\title{
    Solution of the Pierce problem near a corner of a rectilinear flow
    with a polygonal cross-section
}

\author{Igor A. Kotelnikov}
\email{I.A.Kotelnikov@inp.nsk.su}
\affiliation{
    Budker Institute of Nuclear Physics SB RAS,
    Lavrentyev Av. 11, Novosibirsk, 630090, Russian Federation;
}
\affiliation{
    Novosibirsk State University,
    Pirogova Str. 11, Novosibirsk, 630090, Russian Federation
}


\begin{abstract}
    An ill-posed problem of synthesis of the Pierce electrodes for a cylindrical beam with a polygonal cross-section is considered. It is assumed that a beam of charged particles is extracted from a space-charge-limited planar diode and the Pierce electrodes outside of the beam ensure its zero angular divergence. A mathematical statement of the problem presumes a computation of the electrostatic potential outside of the beam that should match the Child-Langmuir 1D potential inside of the beam. An exact solution is first obtained for the potential outside of the beam near its right angle. The solution involves double analytic continuation and a numerical integration of the hypergeometric function and can be used as a benchmark for testing numerical codes. It is shown that equipotential surfaces have fractures that can be pushed away from the corner of the beam by means of smoothing the beam corners. This solution is then generalized to an angle of arbitrary magnitude.
\end{abstract}


\maketitle

\section{Introduction}
\label{s1}


The problem of formation of a cylindrical beam of charged particles with a predefined cross-section arises in development of devices such as multi-aperture particle sources for plasma heating and diagnostics  \cite{Hemsworth+2008RSI_79_02C109, Davydenko+2006RevSciInstr.77.03B902, Davydenko+2007NIM_576_259, Kotelnikov+2008RSI_79_02B702, Sorokin+2010rsi_81, Belchenko+2013aipcp.10.1063.1.4792783}. Such devices compose a cylindrical cutting with a required cross-section from one-dimensional flow of charged particle produced in a planar diode. The particles flow in an idealized endless planar diode is intrinsically rectilinear. It is characterized by zero angular divergence and zero emittance, which means that the particles trajectories do not intersect. Cutting a piece of the flow gives rise to lateral electric field that increases the angular spread of the particles. In many cases, the angular spread should be avoided. According to J.R.~Pierce \cite{Pierce1940, Pierce1954}, the transversal field can be compensated in the entire beam interior by proper shaping of the external electrodes. From the point of view of pure mathematics, the synthesis of such (Pierce) electrodes belongs to a class of ill-posed Cauchy's problems for the Laplace equation \cite{Radley1958.4.125}. It assumes a computation of the electrostatic potential outside of the beam that provides the required compensation of the transversal electric field.

For a particle beam in the form of endless belt, the Pierce problem is two-dimensional. It was solved by Pierce himself. He noted that required solution is given by an analytic continuation of the Child-Langmuir potential $\varphi(z)=z^{4/3}$ \cite{Pierce1940, Pierce1954}, which describes 1D electric field inside a space-charge-limited endless diode. For a cylindrical beam with an arbitrary cross-section, the Pierce problem becomes three-dimensional and the method of analytical continuation fails. A general solution of the Pierce problem in 3D case was found by V.A. Syrovoi \cite{Syrovoi1970PMM_34_4(eng), Syrovoi2012JCTE_57_734} (see also \cite{Syrovoi2004a(eng)}). His algorithm involves elimination of the dependence of the potential on coordinate $z$ along the beam by means of the Laplace transform, two analytical continuations, application of the Riemann method to solve an intermediate 2D problem, and an inverse integral transform using the Lipschitz–Hankel integral.


%

In this note, we apply a slightly modified method of Syrovoi to a beam of charged particles with a cross-section, close to a polygon; in particular, we present an exact solution of the Pierce problem near the right angle of such a beam. This solution is then extended to the case of an arbitrary angle.

The paper is organized as follows. In Section~\ref{s2} we formulate a formal solution of the 3D Pierce problem for a rectilinear beam of charged particles. In Section~\ref{s3} we compute electric potential outside of the beam near the right angle of the beam cross-section. In Section \ref{s4} this solution is generalized to an arbitrary angle. Finally, in Section~\ref{s8} we summarize our results.

\section{A formal solution}
\label{s2}

To begin with, we first formulate the Pierce problem for a cylindrical beam with an arbitrary cross-section. Our goal is to find a solution of the Laplace equation
    \begin{gather}
    \label{1:1}
        \parder{^{2}\varphi}{x^{2}}
        +
        \parder{^{2}\varphi}{y^{2}}
        +
        \parder{^{2}\varphi}{z^{2}}
        =
        0
    \end{gather}
outside of a cylindrical domain assuming the Cauchy boundary conditions
    \begin{gather}
    \label{1:2}
        \varphi=z^{4/3}
        ,
        \qquad
        \parder{\varphi}{n}=0
    \end{gather}
on the side boundary of the cylinder, which is given in a parametric form as
    \begin{gather}
    \label{1:3}
        x=x_{e}(t),
        \qquad
        y=y_{e}(t),
    \end{gather}
where $t$ is a formal parameter, and $\tparder{}{n}$ denotes a derivative, normal to the boundary. The boundary conditions \eqref{1:2} specify the Cauchy problem for the Laplace equation \cite{Lavrentyev1956IANUSSR_20_819(eng)}. Unlike Neumann and Dirichlet problems for elliptic equations (see, eg,  \cite{Koshlyakov+1970(eng)}), it belongs to a class of ill-posed problems of mathematical physics. As the Hadamard example shows \cite{Petrovsky1961(eng)}, a solution to the Cauchy problem for the Laplace equation is unique, but unstable with respect to small perturbations of the boundary conditions.

From the point of view of physics, the space charge of a beam, bounded in transverse directions, inevitably creates a lateral electric field that increases angular divergence of the flux (ie, the emittance of the beam). The lateral field can be compensated in the entire beam interior by a careful shaping of additional electrodes placed outside of the beam. Pierce \cite{Pierce1940, Pierce1954} found the shape of such electrodes in 2D problem by means of analytic continuation of the 1D solution $\varphi(z)=z^{4/3}$ to the complex plane $z+iy$. Pierce's solution
    \begin{gather}
    \label{1:4}
        \varphi(y,z)
        =
        \re(z+iy)^{4/3}
        ,
        \qquad
        y>0,
    \end{gather}
refers to the case of a beam that occupies a half-space $y<0$. In particular, the shape of the electrode with zero potential is determined from the equation
    \begin{gather*}
        \cos\left(
            \tfrac{4}{3}\arctan(y/z)
        \right)
        =
        0
        .
    \end{gather*}
It has the shape of a plane, which forms an angle $ 67.5^{\circ} $ with the plane of the beam boundary.

%
Exact solutions for the beams with circular and elliptical cross-sections were found in Refs.~\cite{Radley1958.4.125, Danilov+1974PMM(eng), Danilov+1974RE(eng), BhattChen2005PhysRevSTAB.8.014201}. Finally, V.A. Syrovoi found a general solution of the Pierce problem for a cylindrical beam with a cross-section of arbitrary shape \cite{Syrovoi1970PMM_34_4(eng)}. However, his work remains unappreciated, and other authors rarely refer to it. Perhaps, this is due to the complexity of Syrovoi's solution. It includes double analytic continuation, and a final expression involves integration of the hypergeometric function whose argument involves analytic continuation of the functions $x_{e}(t)$, $y_{e}(t)$ in a nontrivial way.
In addition, Syrovoi has not presented a clear evidence that the Pierce solution  can be obtained from his formulas. Equivalence of Syrovoi's solutions for the beams with circular and elliptical cross-sections \cite{Danilov+1974PMM(eng), Danilov+1974RE(eng)} to solutions of other authors \cite{Radley1958.4.125, BhattChen2005PhysRevSTAB.8.014201} is not also verified.



In this paper, we use the method of Syrovoi to find the electric potential near one individual right corner of the rectangle, assuming that the beam occupies a quarter of the space, namely
    \begin{gather}
    \label{1:6}
        y < - |x|
        .
    \end{gather}
In this case, a conformal mapping of the exterior of the beam on the complex half-plane is made by a power function, which greatly simplifies the calculations.

\begin{figure}
  \centering
  \includegraphics[width=0.8\columnwidth]{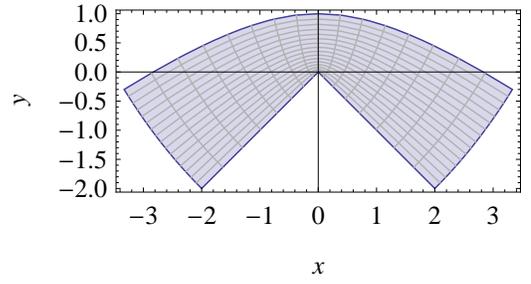}\\
  \caption{
    Mapping of the rectangle $-2<u<2$, $0<v<1$ onto the plane $x,y$.
  }\label{fig:Map}
\end{figure}

Let the cylinder guide $\Gamma$ is defined by parametric equations \eqref{1:3}. It is easy to see that
    \begin{gather}
    \label{2:1}
        x + iy = x_{e}(w)+iy_{e}(w)
    \end{gather}
maps the real axis $v=0$ in the complex plane $w=u+iv$ on $\Gamma$ in the plane $x, y$. Specifically, the transformation
    \begin{gather}
    \label{2:2}
    x_{e}(w) + iy_{e}(w)
    =
    i \left(
        u/i + v
    \right)^{3/2}
    \end{gather}
maps the half-plane $v>0$ on the area immediately outside of the quadrant \eqref{1:6}. The transformation is unique in the plane $w$ with a cut along the negative real half-axis $v<0$; hereinafter we choose the principal value of the power function $t^{n}=\exp(n\ln t)$, which assumes that the argument $t$ is brought to the interval $-\pi <\arg(t) <\pi$ by the modulo $2\pi$. Figure \ref{fig:Map} illustrates the mapping \eqref{2:2}.
Separating the real and imaginary parts in Eq.~\eqref{2:2}, we find
    \begin{gather}
    \label{2:4}
    \begin{aligned}
        x(u,v) &=
        \frac{i}{2}\left[
            \left(
                -iu+v
            \right)^{3/2}
            -
            \left(
                iu+v
            \right)^{3/2}
        \right]
        ,
        \\
        y(u,v) &=
        \frac{1}{2}\left[
            \left(
                -iu+v
            \right)^{3/2}
            +
            \left(
                iu+v
            \right)^{3/2}
        \right]
    .
    \end{aligned}
    \end{gather}
The inverse transformation
    \begin{gather}
    \label{2:6}
    \begin{aligned}
        u(x,y) &=
        \frac{i}{2}
        \left[
            (y-i x)^{2/3}-(y+i x)^{2/3}
        \right]
        ,
        \\
        v(x,y) &=
        \frac{1}{2}
        \left[
            (y-i x)^{2/3}+(y+i x)^{2/3}
        \right]
    \end{aligned}
    \end{gather}
maps the exterior of the quadrant \eqref{1:6} onto the upper half-plane $w$. We will further assume that the beam corner can be smoothed by choosing a coordinate line $v=v_{b}>0$ for the role of the beam boundary. In Fig.~\ref{fig:Map}, the coordinate lines $v=\const$ encircle the right angle, and the coordinate lines $u=\const$ are orthogonal to them.

Eqs.~\eqref{2:4} and~\eqref{2:6} define the first analytic continuation of the two continuations used in Syrovoi's theory. The second analytic continuation is made by replacing $u\to u+i\xi$ in the Laplace equation and the boundary conditions on $\Gamma$. Such a substitution transforms the Laplace equation into a hyperbolic equation that can be solved in a general form by the Riemann method using Laplace transform. We give a final result in a form somewhat different from that obtained by Syrovoi. The electric potential at a point with coordinates $x = x(u, v)$, $y=y(u, v)$ and $z>0 $ is given by the expression
    \begin{gather}
    \label{2:11}
    \varphi
    =
    z^{4/3}
    +
    \frac{1}{18z^{2/3}}
    \int_{{v_{b}}-{v}}^{{v}-{v_{b}}} \,
    {_2F_1}\left(\frac{1}{3},\frac{5}{6};2;-\frac{r^2}{z^2}\right)
    \parder{r^{2}}{v_{b}}
    \,\dif\xi
    ,
    \end{gather}
where ${_2F_1}$ denotes the Gauss hypergeometric function, parameters $u$, $v$ are related to $x$, $y$ by Eq.~\eqref{2:6},
    \begin{multline}
    \label{2:12}
        r^{2}
        =
        \left[
            x(u,v)-x(u+i\xi,v_{b})
        \right]^{2}
        +
        \left[
            y(u,v)-y(u+i\xi,v_{b})
        \right]^{2}
        \\
        =
        \left[
            (i u+v)^{3/2}-(iu+v_{b}-\xi)^{3/2}
        \right]
        \times
        \\
        \left[
            (-iu+v)^{3/2}-(-iu+v_{b}+\xi)^{3/2}
        \right]
        ,
    \end{multline}
and the equations $x=x(u,v_{b})$, $y=y(u,v_{b})$ define a smoothed boundary of the beam.

Since $r^{2}$ is a multi-valued function, a particular branch of $r^{2}$ in the integral \eqref{2:11} should be thoroughly chosen. This is provided by making cuts in the complex plane $u+i\xi$. And what is more important, these cuts should conform to the rules used by computational algorithms when evaluating the power functions such as $(iu+v_{b}-\xi)^{3/2}$. We note that writing down same expression in different forms leads to different results. For example, $(iu+v_{b}-\xi)^{3/2}$ is not the same as $(i)^{3/2}(u+v_{b}/i-\xi/i)^{3/2}$ since these two forms can be evaluated to different complex numbers for the same values of $u$, $v_{b}$, and $\xi$. We carefully tuned the expressions \eqref{2:2}, \eqref{2:4}, and \eqref{2:6} to achieve a satisfactory results so that $r^{2}$ has two cuts in the complex plane $u+i\xi$ along the positive half of axis $\xi$ from $v_{b}$ to $v$ and along the negative half from $-v_{b}$ to $-v$ for any $v$ in the interval of integration over $\xi$ from $v_{b}-v$ to $v-v_{b}$ (recall that $v>v_{b}>0$).

\section{Analysis of the solution}
\label{s3}

\begin{figure*}[!th]
  \centering
  \includegraphics[width=\textwidth]{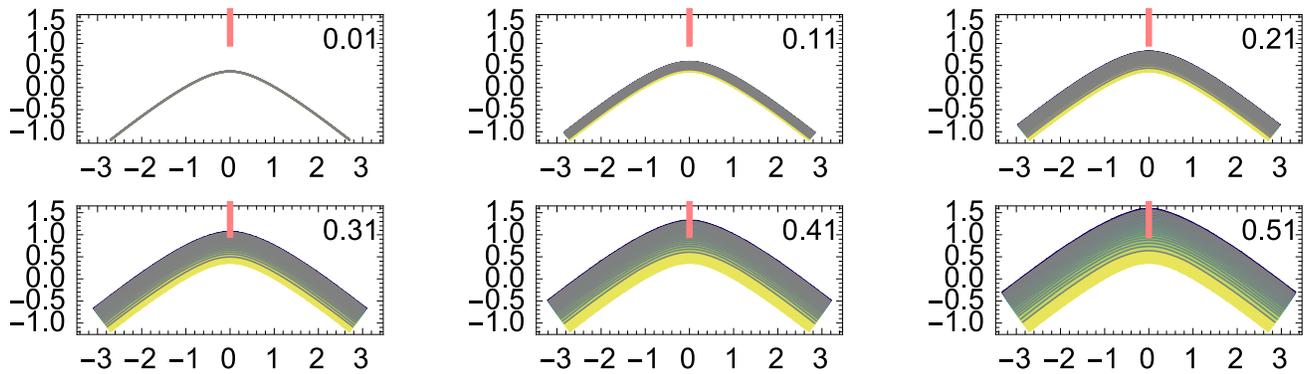}
  \caption{
    (Color online)
    %
    Equipotential lines in the plane $z=\const$ for $ v_ {b} = 0.5 $ and various values of $z$; the values of $z$ are shown in each subfigure, and the equipotential lines are drawn in the interval $0<\varphi<z^{4/3}$; the pink lines indicate the position of the fractures on the equipotential lines.
  }\label{fig:Potential}
\end{figure*}
The result of numerical integration of Eq.~\eqref{2:11} near a smoothed beam corner is shown in Fig.~\ref{fig:Potential} for $v_{b}=0.5$. Pink lines indicate the position of the fractures on the equipotential surfaces. The electric potential $\varphi$ is continuous on the fractures but its normal derivative $\tparder{\varphi}{n}$ jumps; it means that the pink surfaces bear a nonzero surface electric charge. Thus, despite the smoothness of the contour $x=x(u,v_{b})$, $y=y(u,v_{b})$ with $v_ {b}> \varepsilon > 0$ and the conformal mappings \eqref{2:2}, Eq.~\eqref{2:11} defines a discontinuous solution. In the band $v_{b}<v<2v_{b}$ the solution is smooth, but further away from the boundary of the beam the equipotential surfaces experience a kink on the line $x=0$. This phenomenon is formally related to the fact that the imaginary part of the function \eqref{2:12} is discontinuous at $u=0$ if $v>2v_{b}$. From the physical point of view, the presence of the fracture means the impossibility of forming a cylindrical beam with zero emittance and relevant cross-section without introducing charged surfaces, which approach the boundary of the beam as the radius of rounded corners decreases.

Numerical integration of Eq.~\eqref{2:11} for the case $v_{b} \to 0$ yields a quite expected result. The potential $\varphi$ outside the beam is formed by the two 2D Pierce solutions \eqref{1:4} clued at the surface $x=0$.

\section{Arbitrary angle}
\label{s4}

The solution, presented in Sec.~\ref{s2}, can be readily generalized to the case of a beam edge with an angle of arbitrary magnitude. Assuming that the beam occupies the region
    \begin{gather}
    \label{1:6a}
        y < - |x|\cot{\Phi}
        ,
    \end{gather}
we need to replace Eq.~\eqref{2:2} by the generalized transformation
    \begin{gather}
    \label{4:2}
    x_{e}(w) + iy_{e}(w)
    =
    i \left(
        u/i + v
    \right)^{(2\pi-2\Phi)/\pi}
    .
    \end{gather}
Then, Eq.~\eqref{2:4} reads
    \begin{gather}
    \label{4:4}
    \begin{aligned}
        x(u,v) &=
        \frac{i}{2}\left[
            \left(
                -iu+v
            \right)^{(2\pi-2\Phi)/\pi}
            -
            \left(
                iu+v
            \right)^{(2\pi-2\Phi)/\pi}
        \right]
        ,
        \\
        y(u,v) &=
        \frac{1}{2}\left[
            \left(
                -iu+v
            \right)^{(2\pi-2\Phi)/\pi}
            +
            \left(
                iu+v
            \right)^{(2\pi-2\Phi)/\pi}
        \right]
    ,
    \end{aligned}
    \end{gather}
and the inverse transformation \eqref{2:6} becomes
    \begin{gather}
    \label{4:6}
    \begin{aligned}
        u(x,y) &=
        \frac{i}{2}
        \left[
            (y-i x)^{\pi/(2\pi-2\Phi)}-(y+i x)^{\pi/(2\pi-2\Phi)}
        \right]
        ,
        \\
        v(x,y) &=
        \frac{1}{2}
        \left[
            (y-i x)^{\pi/(2\pi-2\Phi)}+(y+i x)^{\pi/(2\pi-2\Phi)}
        \right]
        .
    \end{aligned}
    \end{gather}
When the angle $\Phi$ varies from $0$ (cut angle) to $\pi$ (obtuse angle) the factor $\pi/(2\pi-2\Phi)$ increases from $1/2$ to $\infty$; $\Phi=\pi/4$ corresponds to the right angle. The main expression \eqref{2:11} remains unchanged but Eq.~\eqref{2:12} takes the form
\begin{multline}
    \label{4:12}
        r^{2}
        =
        \left[
            (i u+v)^{(2\pi-2\Phi)/\pi}-(iu+v_{b}-\xi)^{(2\pi-2\Phi)/\pi}
        \right]
        \times
        \\
        \left[
            (-iu+v)^{(2\pi-2\Phi)/\pi}-(-iu+v_{b}+\xi)^{(2\pi-2\Phi)/\pi}
        \right]
        .
    \end{multline}
Main conclusions made in Sec.~\ref{s3} remain valid for arbitrary  $\Phi$. In particular, the fracture appears for any $\Phi\neq \tfrac{1}{2}\pi$, ie., except for the case of planar beam boundary, considered by Pierce.

The Pierce solution \eqref{1:4} for a beam with plane boundary can be recovered from Eqs.~\eqref{2:11} and~\eqref{4:12}. To proof this statement, it is sufficient to take $\Phi=\pi/2$ in Eq.~\eqref{4:12}. Then,
    \begin{gather*}
    \label{2:21}
        x(u,v) = u
        ,\\
        y(u,v) = v
        ,\\
        r^{2} = (v-v_{b})^{2}-\xi^{2}=(y-v_{b})^{2}-\xi^{2}
        ,
    \end{gather*}
and Eq.~\eqref {2:11} takes the form
    \begin{gather*}
    \varphi
    =
    z^{4/3}
    -
    \frac{y}{9z^{2/3}}
    \int_{-y}^{y} \,
    {_2F_1}\left(\frac{1}{3},\frac{5}{6};2;-\frac{y^2-\xi^{2}}{z^2}\right)
    \dif\xi
    ,
    \end{gather*}
if $v_{b} = 0$.
After the substitution $\xi=y\sin(t)$, $\dif\xi=y\cos(t)\dif t$ a computational software program \emph{Mathematica} \cite{WolframMathematica} computes the integral in an analytic form and returns the expression
    \begin{gather*}
        \varphi
        =
        z^{4/3}\left(
            1+y^{2}/z^{2}
        \right)^{2/3}
        \cos\left(
            \tfrac{4}{3}\arctan(y/z)
        \right)
        ,
    \end{gather*}
which is equivalent to Eq.~\eqref{1:4}. This proofs that Syrovoi's theory contains the Pierce solution for a planar beam in the form of endless belt as a particular case.

\section{Discussion}
\label{s8}

%

Our study has confirmed validity of the theory developed by V.A.~Syrovoi,\cite{Syrovoi1970PMM_34_4(eng)} which seems to be not properly evaluated by other researchers. In particular, we have shown that classical 2D Pierce's solution \cite{Pierce1940} can be deduced from that of Syrovoi. We found a solution of the 3D Pierce problem near the sharp or rounded right angle of a rectilinear flow of charged particles. This solution is then generalized to the angle of arbitrary magnitude. Although it demonstrates principal impossibility of constructing the Pierce electrodes because of appearance of fractures in the spatial dependance of the electric potential for a beam with a cross-section that has sharp corners, this solution can be used as a benchmark for existing numerical codes when designing devices with minimized beam emittance. For practical needs, it could also be sufficient to round the corners of the cross-section to achieve a smooth single-valued solution which would mean feasibility of electrodes with Pierce's geometry  in a vicinity of the beam boundary.




\begin{acknowledgments}
    The work was supported by the Ministry of Education and Science of Russian Federation (project RFMEFI61914X0003). The author is grateful to A.A.~Ivanov, V.I.~Davydenko, A.D.~Beklemishev, and \framebox{M.A.~Tiunov} for useful discussions.
\end{acknowledgments}

\bigskip


%

\end{document}